\documentclass[10pt,conference]{IEEEtran}
\usepackage{booktabs} % For formal tables

\usepackage{stmaryrd}
\usepackage{amsmath}
\usepackage{epstopdf}
\usepackage{amssymb}
\usepackage{pifont}
\usepackage[nocompress]{cite}
\usepackage{caption}
\usepackage{graphicx}
\usepackage{array}
\usepackage{listings}
\usepackage[table]{xcolor}
\usepackage{mathpartir}
\usepackage[vlined,ruled,linesnumbered]{algorithm2e}

\definecolor{codegreen}{rgb}{0,0.6,0}
\definecolor{codegray}{rgb}{0.5,0.5,0.5}
\definecolor{codepurple}{rgb}{0.58,0,0.82}
\definecolor{backcolour}{rgb}{0.95,0.95,0.92}
\definecolor{pblue}{rgb}{0.13,0.13,1}
\definecolor{pgreen}{rgb}{0,0.5,0}
\definecolor{pred}{rgb}{0.9,0,0}
\definecolor{pgrey}{rgb}{0.46,0.45,0.48}

 \lstdefinestyle{mystyle2}{
  showspaces=false,
  showtabs=false,
  breaklines=true,
  showstringspaces=false,
  breakatwhitespace=true,
  commentstyle=\color{pgreen},
  keywordstyle=\color{pblue},
  stringstyle=\color{pred},
  basicstyle=\ttfamily,
  moredelim=[il][\textcolor{pgrey}]{$ $},
  moredelim=[is][\textcolor{pgrey}]{\%\%}{\%\%}
}

 \lstdefinestyle{mystyle}{
    backgroundcolor=\color{backcolour},   
    commentstyle=\color{codegreen},
    keywordstyle=\color{magenta}\bfseries,
    numberstyle=\color{codegray},
    stringstyle=\color{codepurple}
}

\begin{document}

%%%%%%%%%%%%%%%%%%%%%%%%%%%%%%%%%%%%%%%%%%%%%%%%%%%%%%%%%%%%%%%%%%%%%%%%%%%%%%%%%%%%%%%%%%%%
% Title
%%%%%%%%%%%%%%%%%%%%%%%%%%%%%%%%%%%%%%%%%%%%%%%%%%%%%%%%%%%%%%%%%%%%%%%%%%%%%%%%%%%%%%%%%%%%

\title{DCert: Find the Leak in Your Pocket}

\author{\IEEEauthorblockN{Mohamed Nassim Seghir}
\IEEEauthorblockA{University College London}}

%\author{\IEEEauthorblockN{M. Nassim Seghir}}
%\IEEEauthorblockA{University College London}
\maketitle
%\IEEEauthorblockA{School of Electrical and\\Computer Engineering\\
%Georgia Institute of Technology\\
%Atlanta, Georgia 30332--0250\\
%Email: http://www.michaelshell.org/contact.html}

%\titlenote{Produces the permission block, and
%  copyright information}
%\subtitle{Extended Abstract}
%\subtitlenote{The full version of the author's guide is available as
%  \texttt{acmart.pdf} document}

%\author{M. Nassim Seghir}
%\authornote{}
%\orcid{}
%\affiliation{%
%  \institution{University College London}
  %\streetaddress{}
  %\city{London} 
  %\postcode{}
%}
%\email{n.seghi@ucl.ac.uk}

%%%%%%%%%%%%%%%%%%%%%%%%%%%%%%%%%%%%%%%%%%%%%%%%%%%%%%%%%%%%%%%%%%%%%%%%%%%%%%%%%%%%%%%%%%%%
% Abstract
%%%%%%%%%%%%%%%%%%%%%%%%%%%%%%%%%%%%%%%%%%%%%%%%%%%%%%%%%%%%%%%%%%%%%%%%%%%%%%%%%%%%%%%%%%%%

\begin{abstract}
Static data-flow analysis has proven its effectiveness in assessing security of applications. One major challenge it faces is scalability to large software. This issue is even exacerbated when additional limitations on computing and storage resources are imposed, as is the case for mobile devices. In such cases the analysis is performed on a conventional computer. This poses two problems. First, a man-in-the-middle attack can tamper with an analyzed application. So once on the mobile device, what guarantees that the actual version is not corrupt. Second, the analysis itself might be broken leading to an erroneous result. As a solution, we present DCert a tool for checking and certifying data-flow properties that consists of two components: a (heavyweight) analyzer and a (lightweight) checker. The analyzer is deployed on a conventional computer. It verifies the conformance of a given application to a specified policy and generates a certificate attesting the validity of the analysis result. It suffices then for the checker, on a mobile device, to perform a linear pass in the application size to validate or refute the certificate as well as the policy. This allows us to separate the verification and the checking process while ensuring a trust relationship between them via the certificate. We describe DCert and report on experimental results obtained for real-world applications.\\
\end{abstract}

%%%%%%%%%%%%%%%%%%%%%%%%%%%%%%%%%%%%%%%%%%%%%%%%%%%%%%%%%%%%%%%%%%%%%%%%%%%%%%%%%%%%%%%%%%%%
% Will be done by appropriate time  
%%%%%%%%%%%%%%%%%%%%%%%%%%%%%%%%%%%%%%%%%%%%%%%%%%%%%%%%%%%%%%%%%%%%%%%%%%%%%%%%%%%%%%%%%%%%


%\begin{CCSXML}
%\end{CCSXML}

%\ccsdesc[500]{Computer systems organization~Embedded systems}

%%%%%%%%%%%%%%%%%%%%%%%%%%%%%%%%%%%%%%%%%%%%%%%%%%%%%%%%%%%%%%%%%%%%%%%%%%%%%%%%%%%%%%%%%%%%

%\keywords{Static analysis, data-flow analysis, certification.}

\maketitle

\section{Introduction}
\label{sec:intro}
Mobile devices are playing a more and more important part in daily life, they are used to perform several tasks and store a variety of information: personal, financial, industrial, etc. Therefore, security and privacy are becoming major concerns. While the security model on Android already provides resource protection via permissions, it also incurs some rigidity on user choices. For example, a photo app obviously requires the \texttt{CAMERA} permission but also might require the \texttt{INTERNET} permission to receive and display some ads. A possible malicious behaviour of that app is to send taken pictures through the Internet without user consent. Android model allows us either to grant or deny permissions which might impede app normal functionality. Ideally, we want to know if there is a link between the \texttt{CAMERA} and the \texttt{INTERNET} permission. Based on that, we decide whether or not to install the app. Static taint analysis permits to check the presence of data-flow paths from the API function for taking pictures (\emph{source}), to the API function for sending files through the Internet (\emph{sink}). It has proven its effectiveness in assessing the security of Android applications \cite{fuchs:cs-tr-4991, flowdroid, WeiROR14, GordonKPGNR15, BarrosJMVDdE15, ErnstJMDPRKBBHVW14, 0029BBKTARBOM15}. 

A main issue with static analysis is scalability. This problem is even exacerbated when additional limitations on computing and storage resources are imposed, as is the case for mobile devices. In case of the previously mentioned tools, the analysis is performed on a conventional computer. This poses two problems. First, a man-in-the-middle attack can tamper with an analyzed application. So once on the mobile device, what guarantees that the actual version is not corrupt. Second, the analysis itself might be broken leading to an erroneous result. 

We present a Proof-Carrying-Code inspired solution~\cite{NeculaL96} that advises the usage of a certificate as an audit for the accountability of the static analysis algorithm. It consists of splitting the analysis process between two parties: a (heavyweight) analyzer and a (lightweight) checker. The analyzer is deployed on a conventional computer. It verifies the conformance of a given application to a specified (data-flow) property and generates a certificate attesting the validity of the analysis result. It suffices then for the checker, on a mobile device, to perform a linear pass in the application size to validate or refute the certificate as well as the property. This allows us to separate the verification and the checking process while ensuring a trust relationship between them via the certificate. We have implemented our approach in a tool called DCert (Droid Certifier) and successfully applied it to real-world applications. In what follows, we describe its main ingredients and report on experimental results.

\section{DCert in Action}
\label{sec:examples}
We illustrate the functionalities of DCert's main ingredients through an example. Consider the simple code in Figure~\ref{fig:run_example} as part of an Android application. To ease the presentation, we omit irrelevant details. We have the root procedure \texttt{foo} which makes call to function \texttt{bar} which, in turn, calls procedures \texttt{getId}, \texttt{Send} and \texttt{getNumber}. Function \texttt{getId} reads the device identifier using the API method \texttt{getDeviceId} at line 4. Similarly, function \texttt{getNumber} returns the number of the actual phone via API method \texttt{getLine1Number} at line 4. Finally, procedure \texttt{Send} is used to send the string it takes as argument as an SMS via API method \texttt{sendTextMessage} at line 5. Both methods \texttt{getDeviceId} and \texttt{getLine1Number} represent sources and \texttt{sendTextMessage} is a sink. We want to verify that the app does not leak information from certain sources to certain sinks and generate a checkable certificate attesting the outcome. For this we have three ingredients. 
\begin{figure*}[t]
\begin{center}
\begin{tabular}{l@{\hspace{0.4in}}c}
\hline
\\
%%%%%%%%%%%%%%%%%%%%%%%%%%%%%%%%%%%%
\begin{minipage}{5cm}	
\begin{lstlisting}[escapechar=\%]
String foo()
{
  String x = bar();
  return x;	
}
\end{lstlisting}
\end{minipage}
%%%%%%%%%%%%%%%%%%%%%%%%%%%%%%%%%%%%%%
    &
%%%%%%%%%%%%%%%%%%%%%%%%%%%%%%%%%%%%
\begin{minipage}{8.3cm}
\begin{lstlisting}[escapechar=\%]
String getId()
{      
  TelephonyManager tm = ...; // get manager
  String x = tm.getDeviceId(); 	
  return x;
}

\end{lstlisting}
\end{minipage}

%%%%%%%%%%%%%%%%%%%%%%%%%%%%%%%%%%%%%%
%%%%%%%%%%%%%%%%%%%%%%%%%%%%%%%%%%%%%%
\\
%%%%%%%%%%%%%%%%%%%%%%%%%%%%%%%%%%%%%%
%%%%%%%%%%%%%%%%%%%%%%%%%%%%%%%%%%%%%%

\begin{minipage}{5cm}	
\begin{lstlisting}[escapechar=\%]
String bar()
{
  String x = getId();	
  Send(x);
  String y = getNumber();
  return y;
}

\end{lstlisting}
\end{minipage}
%%%%%%%%%%%%%%%%%%%%%%%%%%%%%%%%%%%%%%
    &
%%%%%%%%%%%%%%%%%%%%%%%%%%%%%%%%%%%%
\begin{minipage}{8.3cm}
\begin{lstlisting}[escapechar=\%]
String getNumber(String x)
{
  TelephonyManager tm = ...; // get manager
  String x = tm.getLine1Number(); 	
  return x;
}

\end{lstlisting}
\end{minipage}

%%%%%%%%%%%%%%%%%%%%%%%%%%%%%%%%%%%%%%

\\
&
\begin{minipage}{8.3cm}
\begin{lstlisting}[escapechar=\%]
void Send(String x)
{
  String num = "..."; // destination phone number
  SmsManager SM = ...; // get manager
  SM.sendTextMessage(num, null, x, null, null);
}

\end{lstlisting}
\end{minipage}

\\
&
\\
\hline
\end{tabular}
\end{center}
\caption{Simple Java example illustrating potential data flows from sources to sinks. Method \texttt{getDeviceId} is an Android method for obtaining the device identifier, method \texttt{getLine1Number} permits to obtain the phone number and \texttt{sendTextMessage} allows to send text messages (SMS).}
\label{fig:run_example}
\end{figure*}

\subsection{Property}
The first step in our approach is to specify data paths that should not be present in the considered application. Let us use \texttt{id} and \texttt{num} to respectively refer to the sources \texttt{getDeviceId} and \texttt{getLine1Number}. We also write \texttt{sms} for the sink \texttt{sendTextMessage}. We want to express the absence of information leak from sources to sinks in procedure \texttt{foo}. This is done via:
\[
\mathsf{foo\;:\;\neg(\texttt{sms}, \texttt{id}),  \neg(\texttt{sms}, \texttt{num})}
\]
Each pair $(x,y)$ expresses a data-flow from $x$ to $y$ and the symbol $\neg$ represents a negation. Hence, the policy says that there should be no path reading the phone ID and sending it via SMS (in the first pair) and the same applies to the phone number (second pair). If no property is specified, the implicit property (by default) is that no data-flow from any source to any sink is allowed. The next step is to analyse the app to verify the validity of the specified property.      

\subsection{Analyzer}
\begin{figure}
\begin{center}
\includegraphics[scale=.5]{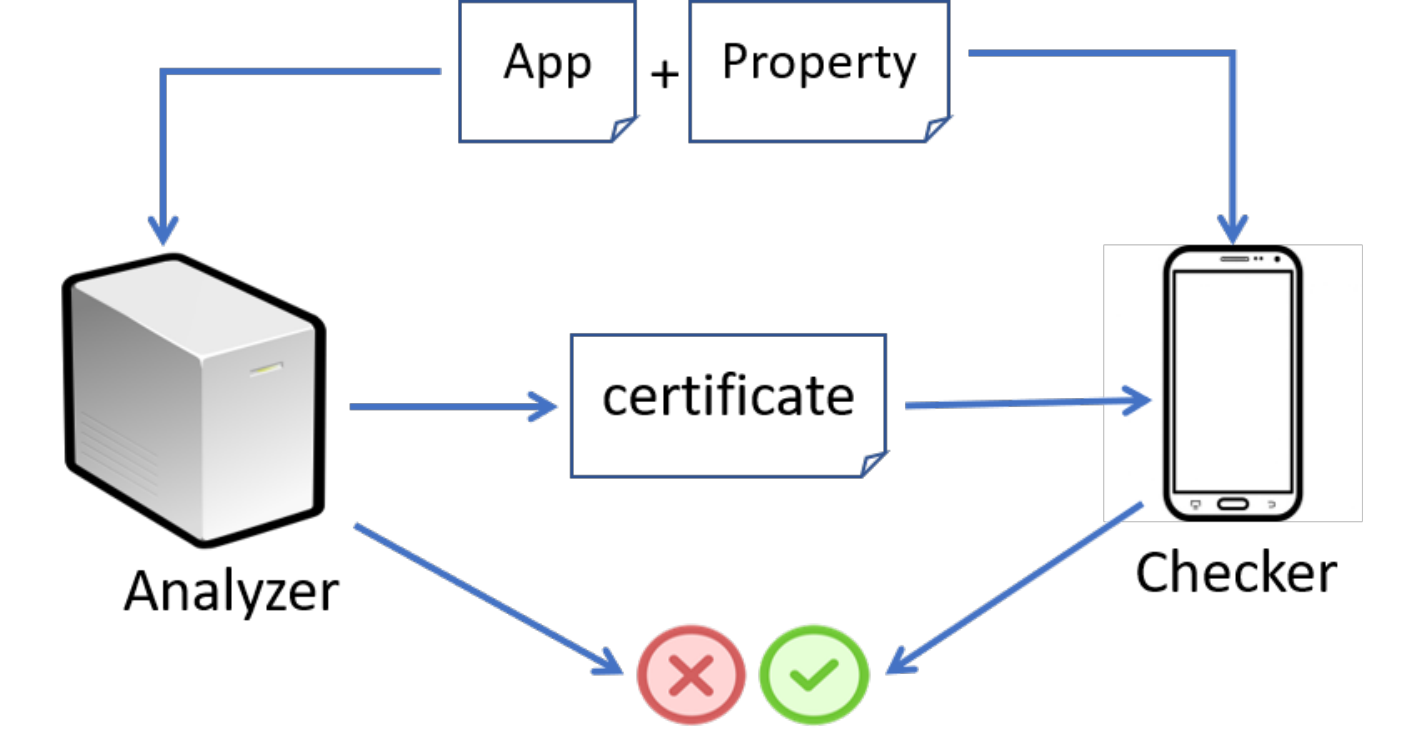}
\caption{DCert main components.}
\label{fig:dcert}
\end{center}
\end{figure}

Figure~\ref{fig:dcert} illustrates the key parts of DCert. The analyzer takes an app and a property as input and answers whether the property is satisfied by the app, and eventually outputs a certificate corroborating the outcome of the analysis. The analyzer implements an inter-procedural data-flow analysis. For each function, it computes a summary which consists of a set of pairs $(x,y)$ expressing the existence of a data-flow from $y$ to $x$. A summary of a given procedure only includes elements visible outside of it. Hence, local variables will not appear in a summary. During the analysis, when a function is invoked from another one, its summary is used instead of re-analysing it. This process is iterated until a fix-point is reached. For illustration, consider Figure~\ref{fig:summar_iter}. It shows the summary computed for the different methods of our previous example (Figure~\ref{fig:run_example}) at each iteration. Initially (iteration 0), all function summaries are empty.   

After iteration 1, empty summaries are still associated with procedures \texttt{foo} and \texttt{bar}, however, summaries for procedures \texttt{getId}, \texttt{getNumber} and \texttt{Send} are updated. The symbol $\mathsf{ret}$ models the return value of a method. Hence, summaries $\mathsf{(ret,id)}$ and $\mathsf{(ret,num)}$, respectively, express flows of the phone identifier (in \texttt{getId}) and the phone number (in \texttt{getNumber}) to a return statement. Similarly, $\mathsf{(sms, x)}$ expresses the presence of a data-flow from the argument $\mathsf{x}$ of procedure $\mathsf{Send}$ to the sink $\mathsf{sms}$. 

After iteration 2, the summary for procedure $\mathsf{bar}$ is updated as summaries associated with its callees changed in the previous step (iteration 1), meaning potential new data paths. For example, $\mathsf{(sms, id)}$  is due to the path $\mathsf{Send} \leftarrow \mathsf{x} \leftarrow \mathsf{getId}$ in procedure $\mathsf{bar}$, when procedures $\mathsf{Send}$ and $\mathsf{getId}$ are substituted with their summaries. 

Finally, the last iteration (3) updates the summary for $\mathsf{foo}$ by just propagating $\mathsf{bar}$'s summary. At this state a fix-point is reached and no further changes will be induced. The presence of $\mathsf{(sms, id)}$ in the summary associated with $\mathsf{foo}$ implies the violation of one rule, namely $\mathsf{\neg(sms, id)}$, but the other rule, $\mathsf{\neg(sms, num)}$, is not violated.

The final map (iteration 3)  represents the certificate. It will be returned by the analyser together with a report indicating rules from the policy that are violated.

\begin{figure}
\footnotesize
{
\begin{tabular}{@{\hspace{-0.4in}}r c|c|c|c}
%@{\hspace{0.4in}}c{\hspace{0.4in}}c{\hspace{0.4in}}c{\hspace{0.4in}}c}
%&&&&
\cline{2-5}
&	\multicolumn{4}{c}{Iteration} \\
\cline{2-5}
 & {\bf 0} &{\bf 1} &{\bf 2} & {\bf 3} \\
\cline{2-5}
\
$\mathsf{foo}$: & - & - & - & $\mathsf{(sms, id)}, \mathsf{(ret, num)}$\\
$\mathsf{bar}$: & - & - & $\mathsf{(sms, id)}, \mathsf{(ret, num)}$ & $\mathsf{(sms, id)}, \mathsf{(ret, num)}$\\
$\mathsf{getId}$: & - &$\mathsf{(ret, id)}$ & $\mathsf{(ret, id)}$ & $\mathsf{(ret, id)}$\\
%$\mathsf{getNumber}$: &$\mathsf{(ret, num)$ & &\\
$\mathsf{getNumber}$: & - &$\mathsf{(ret, num)}$ &  $\mathsf{(ret, num)}$ & $\mathsf{(ret, num)}$\\
$\mathsf{Send}$: & - & $\mathsf{(sms, x)}$ & $\mathsf{(sms, x)}$ & $\mathsf{(sms, x)}$ \\
\cline{2-5}
\end{tabular}
}
\caption{Iterative computation of function summaries. A pair $(x,y)$ models a data-flow from $y$ to $x$.}
\label{fig:summar_iter}
\end{figure}

\subsection{Checker}
Now the question is how can a client of the analysis trust its claim? The analysis might contain errors or, even worst, an attacker can claim app safety without applying the analysis at all. For this, the computed map will serve as a certificate. To test its validity, we just need to locally check that the summary of each method is valid by assuming the validity of the summaries of its callees. For example, assuming the summary for $\mathsf{bar}$ is $\{\mathsf{(sms, id)}, \mathsf{(ret, num)}\}$, the summary for $\mathsf{foo}$ must be $\{\mathsf{(sms, id)}, \mathsf{(ret, num)}\}$, which is the case, otherwise we have an inconsistency. This is performed by the checker which takes as input a certificate (computed map) a property and an app, and answers whether the certificate is valid. In addition, if the certificate is valid, it also verifies whether the certificate entails the property. 

Certificate checking is lighter than certificate generation as we do not need to compute a fix-point. Instead, it is performed in a single pass. It has a linear complexity in the number of map entries (functions) and a constant space complexity as we just perform checks without generating information that need to be stored.

\subsection{About Certificate Resilience} Let us test the resilience of the previously generated  certificate with respect to possible tampering scenarios. 

We first omit $\mathsf{(sms,id)}$ from the entries corresponding to both \textsf{foo} and \textsf{bar}. While this will not be detected when checking \textsf{foo} as consistency is not broken, it will be detected when checking \textsf{bar}. Indeed, consistency of the certificate implies that $\mathsf{(sms,id)}$ must be associated with $\mathsf{foo}$ as a consequence of $\mathsf{(ret,id)}$ and $\mathsf{(sms, x)}$ associated with \textsf{getId} and \textsf{Send} respectively. 

Let us have a more extreme scenario where we remove $\mathsf{(-,id)}$ from all entries. This case is also detected as \textsf{id} refers to the API method \textsf{getDeviceId} which represents a seed for the analysis. Hence, consistency implies that $\mathsf{(x,id)}$ must be associated with \textsf{getId} as it directly invokes \textsf{getDeviceId}. 

How about suppressing all the entries? Our analysis will detect this case as the first step in the certificate integrity check is to make sure that all methods used in the program have entries in the map. 

Now, let us modify the code but preserve the certificate as it is, sot that it does not exhibit the changes. Assume we add the call \texttt{Send(y)} just before the return statement at line 6 in procedure \texttt{bar} (Figure~\ref{fig:run_example}). While the checker does not detect inconsistencies with respect to the entry of procedure \texttt{foo}, it finds out that the certificate is no longer valid with respect to the entry of \texttt{bar}. Indeed, $\mathsf{(sms,num)}$ should have been present in \texttt{bar}'s entry.

\iffalse
\subsection{Implicit Dependencies}
We use a well-established approach proposed by Ferrante et al~\cite{FerranteOW87} to compute control dependencies. Based on that, let us have function $\mathsf{control\_dep}(\mathit{CFG}, \ell , M)$ that takes as parameters a control flow graph $CFG$, a location $\ell$ and a map $M$ associating locations with sets of facts. As result, it returns the set of facts induced by control dependencies for location $\ell$. We make call to $\mathsf{control\_dep}$ in the next section when presenting our analysis algorithm. 

\subsection{Handling Unavailable Code}
One challenge we have faced is taking into account library calls. As the code is often unavailable, we need to over-approximate the effect of library APIs on program variables. Our solution is similar to the one adopted by Flowdroid \cite{Flowdroid}: we use two rules to model the effect of library calls. The first rule assumes that the result of a method depends on its parameters as well as the receiver object. The second rule assumes that the receiver object depends the method parameters. For example, for a method that appends a character to a string, we have a rule modelling that the result depends on the appended character.     
\fi

\section{Implementation and Experiments}
\label{sec:experiments}
DCert is written in Python and uses Androguard\footnote{https://github.com/androguard} as front-end for parsing and decompiling Android applications. It accepts Android applications in bytecode format (APK), so it does not require source code. One can simply download an app from a store of choice and analyse it. As mentioned previously, DCert has two main components: Analyser and Checker.   

The analyser runs on a conventional desktop computer. It takes as input an app and a property and returns, as output, a report about the analysis result together with a certificate. Concretely speaking, the property consists of a file containing all sources and sinks that should be taken into account. 

The checker accepts an application, a certificate and a property as input, and answers whether the certificate is valid with respect to the application taken as input. If the certificate is valid, it also reports on whether the property is fulfilled. The checker can run either on a mobile device or a conventional machine. The mobile device version is provided as An android app. As it is written is Python, we use kivy\footnote{https://kivy.org} to facilitate its deployment.

\paragraph*{\bf Real-world Applications.}
We were able to successfully apply DCert to two popular and largely used applications: \textsf{Facebook} and \textsf{FacebookMessenger}. We downloaded both directly from the Google Play store\footnote{https://play.google.com/store/apps}. The two apps are of decent size with each of them containing more than 6000 methods.

In our experiments, we used a typical Linux desktop to run the verifier and a Samsung J5 mobile phone (  Exynox7870 Octa 1.6GHz processor, 2GB RAM), running Android, to host the checker. We analysed both apps with respect the default property, i.e., find all data leaks between any source and any sink.   

First, we call the analyser to verify the validity of the property and to generate a certificate. In a second step, the checker is invoked to check the generated certificate. As a result, no leaks where found in both apps, and the checker was able to confirm the validity of the generated certificate. The whole running time for the checker is around 39 minutes for the \textsf{Facebook} app and 44 minutes for \textsf{FacebookMessenger}. While runtime difference is huge between running the checker on a mobile device and on a desktop computer, these results are encouraging given the size of the considered applications and the limitations of mobile devices. Moreover, we identified regions in our code and in the used front-end that we can optimise to reduce execution time.

\section{Related Work} 
\label{sec:related}
Android security is an active area of investigation, many tools for analyzing security aspects of Android have emerged. Some rely on dynamic analysis \cite{aurasium, EnckGCCJMS10,reina:copperdroid,ZhangYXYGNWZ13,BackesGHMS13}. Other tools are based on static analysis \cite{flowdroid, scandal, FahlHMSBF12,AuZHL12,ChinFGW11, ChenJDDMMWRS13, JeonMVFRFM12}. We are interested in the last category (static analysis) as our aim is to certify the absence of bad behaviors. Our work is a complement to these tools as we are not only interested in analyzing applications, but also to return a verifiable certificate attesting the validity of the analysis result.

The idea of associating proofs with code was initially proposed by Necula under the moniker \emph{Proof-Carrying Code} (PCC) \cite{NeculaL96, Necula97}. It was then used to support resource policies for mobile code~\cite{AspinallM05,BartheCGJP07}. Furthermore, Desmet et al. presented an implementation of PCC for the .NET platform \cite{DesmetJMPPSV08}. While the tool EviCheck \cite{SeghirA15, Seghir0M16} is also based on a similar idea and targets Android, it is unable to analyse data-flow properties. Cassandra also applies PCC to Android \cite{LortzMSBSW14}. Their approach proposes a type system to precisely track information flows. While precision is an advantage, it is hard to assess the practicability of their approach as no experiments involving real-world applications are reported. Our approach is applicable to real-world large applications.      
\section{Conclusion and Further Work}
\label{sec:conclusion}
We presented DCert a tool for analysing and certifying data-flows, inspired by the \emph{proof-carrying-code} paradigm. DCert allows to analyse applications on mobile devices by splitting the certification process between two components: a (heavyweight) analyzer and a (lightweight) checker, and using a certificate as a mechanism for ensuring trust between the two parties. We described DCert's implementation and reported on experimental results obtained for real-world applications. DCert represents an outlook for future app stores, where apps are equipped with contracts (policies) and checkable evidence (certificate). We are not aware of any other tool that implements a similar certification scheme and is scalable to real-world applications.

\bibliographystyle{abbrv}
\bibliography{biblio}

\begin{thebibliography}{10}

\bibitem{flowdroid}
S.~Arzt, S.~Rasthofer, C.~Fritz, E.~Bodden, A.~Bartel, J.~Klein, Y.~L. Traon,
  D.~Octeau, and P.~McDaniel.
\newblock Flowdroid: precise context, flow, field, object-sensitive and
  lifecycle-aware taint analysis for android apps.
\newblock In {\em {PLDI}}, pages 259--269, 2014.

\bibitem{AspinallM05}
D.~Aspinall and K.~MacKenzie.
\newblock Mobile resource guarantees and policies.
\newblock In {\em CASSIS}, pages 16--36, 2005.

\bibitem{AuZHL12}
K.~W.~Y. Au, Y.~F. Zhou, Z.~Huang, and D.~Lie.
\newblock Pscout: analyzing the {Android} permission specification.
\newblock In {\em ACM Conference on Computer and Communications Security},
  pages 217--228, 2012.

\bibitem{BackesGHMS13}
M.~Backes, S.~Gerling, C.~Hammer, M.~Maffei, and P.~von Styp-Rekowsky.
\newblock Appguard - enforcing user requirements on {Android} apps.
\newblock In {\em TACAS}, pages 543--548, 2013.

\bibitem{BarrosJMVDdE15}
P.~Barros, R.~Just, S.~Millstein, P.~Vines, W.~Dietl, M.~d'Amorim, and M.~D.
  Ernst.
\newblock Static analysis of implicit control flow: Resolving java reflection
  and android intents {(T)}.
\newblock In {\em {ASE}}, pages 669--679, 2015.

\bibitem{BartheCGJP07}
G.~Barthe, P.~Cr{\'e}gut, B.~Gr{\'e}goire, T.~P. Jensen, and D.~Pichardie.
\newblock The mobius proof carrying code infrastructure.
\newblock In {\em FMCO}, pages 1--24, 2007.

\bibitem{ChenJDDMMWRS13}
K.~Z. Chen, N.~M. Johnson, V.~D'Silva, S.~Dai, K.~MacNamara, T.~Magrino, E.~X.
  Wu, M.~Rinard, and D.~X. Song.
\newblock Contextual policy enforcement in {Android} applications with
  permission event graphs.
\newblock In {\em NDSS}, 2013.

\bibitem{ChinFGW11}
E.~Chin, A.~P. Felt, K.~Greenwood, and D.~Wagner.
\newblock Analyzing inter-application communication in {Android}.
\newblock In {\em MobiSys}, pages 239--252, 2011.

\bibitem{DesmetJMPPSV08}
L.~Desmet, W.~Joosen, F.~Massacci, P.~Philippaerts, F.~Piessens, I.~Siahaan,
  and D.~Vanoverberghe.
\newblock Security-by-contract on the .net platform.
\newblock {\em Inf. Sec. Techn. Report}, 13(1):25--32, 2008.

\bibitem{EnckGCCJMS10}
W.~Enck, P.~Gilbert, B.~gon Chun, L.~P. Cox, J.~Jung, P.~McDaniel, and
  A.~Sheth.
\newblock Taintdroid: An information-flow tracking system for realtime privacy
  monitoring on smartphones.
\newblock In {\em OSDI}, pages 393--407, 2010.

\bibitem{ErnstJMDPRKBBHVW14}
M.~D. Ernst, R.~Just, S.~Millstein, W.~Dietl, S.~Pernsteiner, F.~Roesner,
  K.~Koscher, P.~Barros, R.~Bhoraskar, S.~Han, P.~Vines, and E.~X. Wu.
\newblock Collaborative verification of information flow for a high-assurance
  app store.
\newblock In {\em {CCS}}, pages 1092--1104, 2014.

\bibitem{FahlHMSBF12}
S.~Fahl, M.~Harbach, T.~Muders, M.~Smith, L.~Baumg{\"a}rtner, and
  B.~Freisleben.
\newblock Why {Eve} and {Mallory} love {Android}: an analysis of {Android}
  {SSL} (in)security.
\newblock In {\em {CCS}}, pages 50--61, 2012.

\bibitem{fuchs:cs-tr-4991}
A.~P. Fuchs, A.~Chaudhuri, and J.~S. Foster.
\newblock {SCanDroid: Automated Security Certification of Android
  Applications}.
\newblock Technical Report CS-TR-4991, Department of Computer Science,
  University of Maryland, College Park, November 2009.

\bibitem{GordonKPGNR15}
M.~I. Gordon, D.~Kim, J.~H. Perkins, L.~Gilham, N.~Nguyen, and M.~C. Rinard.
\newblock Information flow analysis of android applications in droidsafe.
\newblock In {\em {NDSS}}, 2015.

\bibitem{JeonMVFRFM12}
J.~Jeon, K.~K. Micinski, J.~A. Vaughan, A.~Fogel, N.~Reddy, J.~S. Foster, and
  T.~D. Millstein.
\newblock Dr. {Android} and mr. hide: fine-grained permissions in {Android}
  applications.
\newblock In {\em SPSM@CCS}, pages 3--14, 2012.

\bibitem{scandal}
J.~Kim, Y.~Yoon, K.~Yi, and J.~Shin.
\newblock Scandal: Static analyzer for detecting privacy leaks in {Android}
  applications.
\newblock In {\em (MOST)}, 2012.

\bibitem{0029BBKTARBOM15}
L.~Li, A.~Bartel, T.~F. Bissyand{\'{e}}, J.~Klein, Y.~L. Traon, S.~Arzt,
  S.~Rasthofer, E.~Bodden, D.~Octeau, and P.~D. McDaniel.
\newblock Iccta: Detecting inter-component privacy leaks in android apps.
\newblock In {\em {ICSE}}, pages 280--291, 2015.

\bibitem{LortzMSBSW14}
S.~Lortz, H.~Mantel, A.~Starostin, T.~B{\"{a}}hr, D.~Schneider, and A.~Weber.
\newblock Cassandra: Towards a certifying app store for android.
\newblock In {\em {SPSM@CCS}}, pages 93--104, 2014.

\bibitem{Necula97}
G.~C. Necula.
\newblock Proof-carrying code.
\newblock In {\em POPL}, pages 106--119, 1997.

\bibitem{NeculaL96}
G.~C. Necula and P.~Lee.
\newblock Safe kernel extensions without run-time checking.
\newblock In {\em OSDI}, pages 229--243, 1996.

\bibitem{reina:copperdroid}
A.~Reina, A.~Fattori, and L.~Cavallaro.
\newblock A system call-centric analysis and stimulation technique to
  automatically reconstruct {Android} malware behaviors.
\newblock In {\em Proceedings of the 6th European Workshop on System Security
  (EUROSEC 2013)}, 2013.

\bibitem{SeghirA15}
M.~N. Seghir and D.~Aspinall.
\newblock Evicheck: Digital evidence for android.
\newblock In {\em {ATVA}}, pages 221--227, 2015.

\bibitem{Seghir0M16}
M.~N. Seghir, D.~Aspinall, and L.~Marekova.
\newblock Certified lightweight contextual policies for android.
\newblock In {\em {S}ec{D}ev, 3-4, 2016}, pages 94--100, 2016.

\bibitem{WeiROR14}
F.~Wei, S.~Roy, X.~Ou, and Robby.
\newblock Amandroid: {A} precise and general inter-component data flow analysis
  framework for security vetting of android apps.
\newblock In {\em {CCS}}, pages 1329--1341, 2014.

\bibitem{aurasium}
R.~Xu, H.~Sa{\"\i}di, and R.~Anderson.
\newblock Aurasium: Practical policy enforcement for {Android} applications.
\newblock In {\em {USENIX} Security Symposium}, pages 539--552, 2012.

\bibitem{ZhangYXYGNWZ13}
Y.~Zhang, M.~Yang, B.~Xu, Z.~Yang, G.~Gu, P.~Ning, X.~S. Wang, and B.~Zang.
\newblock Vetting undesirable behaviors in android apps with permission use
  analysis.
\newblock In {\em {CCS}}, pages 611--622, 2013.

\end{thebibliography}
\end{document}